\documentclass[journal]{IEEEtran}

\usepackage{cite}
\usepackage{graphicx}
\usepackage{amsmath}
\usepackage{array}
\usepackage{multirow}
\usepackage{caption}
\usepackage{amssymb}
\usepackage{longtable}
\usepackage{color}
\usepackage{makecell}
\usepackage{booktabs}
\usepackage{algorithmic}
\usepackage[ruled, linesnumbered]{algorithm2e}
\usepackage{multirow}
\usepackage[table,xcdraw]{xcolor}

\begin{document}

\title{WUDI: A Human Involved Self-Adaptive Framework to Prevent Childhood Obesity in Internet of Things Environment}

\author{Euijong Lee, Jaemin Jung, Gee-Myung Moon, Seong-Whan Lee*, and Ji-Hoon Jeong*
\thanks{
\textit{(*Corresponding author: Seong-Whan Lee and Ji-Hoon Jeong)}}
\thanks{Euijong Lee, Jaemin Jeong, and Ji-Hoon Jeong with School of Computer Science, Chungbuk National University, Cheongju-si 28644, South Korea (e-mail: kongjjagae@cbnu.ac.kr, jmjung@cbnu.ac.kr, and jh.jeong@chungbuk.ac.kr)}
\thanks{Gee-Myung Moon with the Injewelme corporation, Seoul 02841, South Korea. E-mail: gmmoon@injewelme.com}
\thanks{Seong-Whan Lee is affiliated with the Department of Artificial Intelligence, Korea University, Seongbuk-gu, Seoul 02841, South Korea. E-mail: sw.lee@korea.ac.kr.}
}

\markboth{}%
{}

\maketitle

\begin{abstract}

The Internet of Things (IoT) connects people, devices, and information resources, in various domains to improve efficiency. The healthcare domain has been transformed by the integration of the IoT, leading to the development of digital healthcare solutions such as health monitoring, emergency detection, and remote operation. This integration has led to an increase in the health data collected from a variety of IoT sources. Consequently, advanced technologies are required to analyze health data, and artificial intelligence has been employed to extract meaningful insights from the data. Childhood overweight and obesity have emerged as some of the most serious global public health challenges, as they can lead to a variety of health-related problems and the early development of chronic diseases. To address this, a self-adaptive framework is proposed to prevent childhood obesity by using lifelog data from IoT environments, with human involvement being an important consideration in the framework. The framework uses an ensemble-based learning model to predict obesity using the lifelog data. Empirical experiments using lifelog data from smartphone applications were conducted to validate the effectiveness of human involvement and obesity prediction. The results demonstrated the efficiency of the proposed framework with human involvement in obesity prediction. The proposed framework can be applied in real-world healthcare services for childhood obesity.

\end{abstract}


\IEEEpeerreviewmaketitle

\section{Introduction} \label{sec_intro}

\IEEEPARstart{T}{he} Internet of Things (IoT) interconnects various entities having a distinct existence including physical or non-physical (e.g., sensor, actuator, digital user, and human user). With various advantages of interconnection, IoT is applied to various fields such as smart farming, automobiles, smart homes, and smart factories. Especially, the healthcare field has been revolutionized by the integration of IoT technologies, leading to the development of connected health, which refers to any digital healthcare solution such as remote operation, continuous health monitoring, and emergency detection. The application of IoT in the healthcare field helps to improve the quality of personalized healthcare because traditional healthcare is unable to accommodate everyone’s requirements \cite{sundaravadivel2017everything, yu2021wifi}. The increasing volume of health data generated from various sources, such as electronic health records, wearable devices, and mobile applications, has created the need for advanced technologies to process and analyze the data. This has led to the integration of artificial intelligence (AI) to extract meaningful insights from health data and provide personalized healthcare services \cite{abualsaud2022machine, yu2018artificial, eldele2022adast, sworna2021towards}. Various studies have applied AI to IoT in healthcare, including COVID-19 identification, diabetes prediction, fall detection, remote healthcare monitoring, and heart disease diagnosis \cite{alshamrani2022iot, sworna2021towards, eldele2022adast}. 

Childhood overweight and obesity have increased over the last few decades globally \cite{abarca2017worldwide}, and they can cause early development of diabetes, cardiovascular diseases, fatty liver, and precocious puberty \cite{caprio2020childhood, ahmed2009childhood}. Therefore, this is one of the most serious global public health challenges \cite{caprio2020childhood}, and healthcare service is required to prevent child obesity. 

We propose a self-adaptive framework for personalized child healthcare to prevent obesity using IoT environments. Child lifestyles are diverse and dynamic; thus, adaptive healthcare services are required for each child. Self-adaptive software satisfies the requirements in dynamic environments; thus, self-adaptive software is applicable to healthcare services for children. Such services also need to encourage the participation of children; thus, effective human involvement is inevitable \cite{nunes2015survey}. Therefore, the proposed framework focused on human involvement by using monitoring, analysis, planning, and executing (MAPE) loop, which is a prominent control loop, to develop self-adaptive software. Rewards concepts were implemented to incentivize human participation within the loop. Furthermore, the proposed framework incorporated an ensemble-based learning model to predict obesity based on the lifelog data of children. Empirical experiments were conducted using smartphone applications to evaluate the effectiveness of reward-based human involvement and obesity prediction. The results demonstrated that appropriate rewards could motivate children and reduce body mass index (BMI) which is an obesity metric. The prediction results were reasonable, implying that the proposed approach can be applied in healthcare services to address childhood obesity.

The remainder of this article is organized as follows. Section \ref{sec_backGroundRelatedWork} provides background and related work. Section \ref{sec_wudi} introduces the proposed IoT self-adaptive framework for child healthcare. Section \ref{sec_empiricalEvaluation} presents the results of the empirical experiments. Section \ref{sec_discussion} discusses the limitation and future work, and Section \ref{sec_conclusion} concludes this paper. 
\section {background and related work} \label{sec_backGroundRelatedWork}
In this section, the background and related work are described. Self-adaptive software is introduced in Section \ref{sec_selfAdaptiveSoftware}. Section \ref{sec_healthcareDomain} briefly gives an overview of ICT technologies applied in healthcare fields, and IoT and AI-related healthcare studies for childhood obesity are introduced.

\subsection{Self-Adaptive Software} \label{sec_selfAdaptiveSoftware}

Self-adaptive software aims to autonomously adjust its various attributes in response to the detected context \cite{salehie2009self}. Context refers to various environmental factors that can potentially affect the software. Self-adaptive software requires continuous monitoring and adaptation to the context changes; thus, an adaptation cycle is required, and the cycle is called MAPE loop. MAPE loop is generally applied in self-adaptive software and consists of four processes \cite{salehie2009self}:

\begin{itemize}
    \item Monitoring, which is responsible for collecting data from the software and operating environment.
    \item Analyzing (detecting), which is responsible for analyzing the symptoms using data from the monitoring process. 
    \item Planning (deciding), which is responsible for deciding how to change artifacts or attributes to improve performance. 
    \item Executing (acting), which is responsible for applying changes. 
\end{itemize}

The concept of self-adaptive software is applied in various fields such as healthcare \cite{serhani2020self}, smart farming \cite{lee2019self}, smart homes\cite{lee2022self}, and smart buildings \cite{andrade2021multifaceted}. In this study, it is assumed that health conditions of children can be continuously changed and require adaptation to respond to these health changes. Therefore, the concept of self-adaptive software was applied to manage childhood healthcare.
\subsection{AI and IoT Applied in Healthcare Domain} \label{sec_healthcareDomain}

The integration of AI and IoT is revolutionizing healthcare \cite{yu2018artificial, islam2022internet}. In several studies, IoT sensors including medical devices (e.g., force, heart rate, pulse oximetry, pressure, glucose, temperature, electromyogram, electrocardiogram, and electroencephalogram), smartphones (e.g., gyroscope, accelerometer, and microphone), and wearable devices were developed and used to collect healthcare-related data via the Internet \cite{sworna2021towards, islam2022internet}. Various AI algorithms were also applied to address several healthcare applications using the collected data such as remote monitoring, chronic disease detection, medication management, telehealth, telesurgery, homecare, and early care \cite{alshamrani2022iot, islam2022internet}. Previous research indicates that the integration of AI and IoT in healthcare can be extended to various health-related applications. However, there are critical factors in integrating AI and IoT in the healthcare domain: data collection plan and application of appropriate AI algorithms. In the data collection plan, it is essential to determine which types of data can be collected, and whether the collected data can provide insights into specific healthcare issues. Naturally, the data plan includes how to collect data sufficient for the application of AI algorithms. Moreover, in the application of AI, the appropriate AI algorithms are selected to solve healthcare issues using the collected IoT data. 

This study focused on childhood obesity; therefore, recent studies that focused on childhood obesity-related healthcare using AI and IoT are discussed. Gupta et al. \cite{gupta2022obesity} proposed childhood obesity prediction using electronic health records (EHR) that contained several features such as weight, heart rate, age, temperature, blood pressure, and congenital disease; PEDSnet dataset \cite{pedsnet} was used for the obesity prediction. The dataset contained patient records including at least 5 years of medical history and no evidence of specific disease (i.e., type 1 diabetes, cancer, sickle cell disease, developmental delay, or other complex medical conditions). The experimental results showed efficient prediction results in long-term prediction from 1 to 3 years in advance. Pang et al. \cite{pang2021prediction} applied seven machine-learning models (i.e., decision tree (DT), support vector machine (SVM), linear regression, neural network, Gaussian naïve Bayes, Bernoulli naïve Bayes, and extreme gradient boosting (XGB)) to predict future childhood obesity at the age of 2 years. EHR data from patients were used to generate the prediction models, and the performance results showed that machine-learning models can effectively predict childhood obesity. 

Some studies applied IoT in preventing childhood obesity. An IoT-based microservice platform called OCARIoT was proposed to prevent obesity and promote childhood healthy habits \cite{bastida2023promoting, de2020agent}. In OCARIoT, data were acquired using IoT devices, and the dataset was applied for personalized healthcare monitoring. A rule-based approach was applied to support decision-making and personalized health coaching plans. A prototype of OCARIoT was implemented and the results demonstrated that the ecosystem could assess the behaviors of children and help them to achieve healthy habits. Yacef et al. \cite{yacef2018supporting} proposed an educational tool named iEngage, which collected physical activities using wearable devices and provided objective feedback to help learning activities. A pilot study was conducted and the results demonstrated that the tool could be applied to change positive behaviors. Verma et al. \cite{verma2018cloud} proposed an IoT-based framework to predict potential diseases, including childhood obesity. The framework assumed that various attributes, such as blood pressure, temperature, and family history, could be collected. For experiments, a UCI data repository \cite{asuncion2007uci} was used, and machine-learning algorithms (i.e., kNN, SVM, DT, and NB) were applied. The results were efficient in disease predictions including obesity.

This study focused on childhood obesity; thus, obesity-related data were identified and collected from IoT environments. An ensemble-based learning model was applied to predict childhood obesity using the collected data. To integrate data collection and prediction, a comprehensive framework with a self-adaptive concept was proposed, and the details are described in Section \ref{sec_wudi}.

\section{WUDI: self-adaptive framework for personalized child healthcare} \label{sec_wudi}

A self-adaptive AI-based obesity prediction framework named ``Would You Do It?'' (WUDI) is proposed for child healthcare. In this section, the proposed framework is introduced. Section \ref{sec_IoTArchitecture} describes WUDI with IoT architectural view. The framework with a self-adaptive concept and human-in-the-loop mechanism is described in Section \ref{sec_selfAdaptiveFramework}. The AI-based obesity prediction model is introduced in Section \ref{sec_ensembleModel}.

\subsection{IoT Architectural Overview} \label{sec_IoTArchitecture}

\begin{figure}[]
\centering
\includegraphics[width=6.8cm]{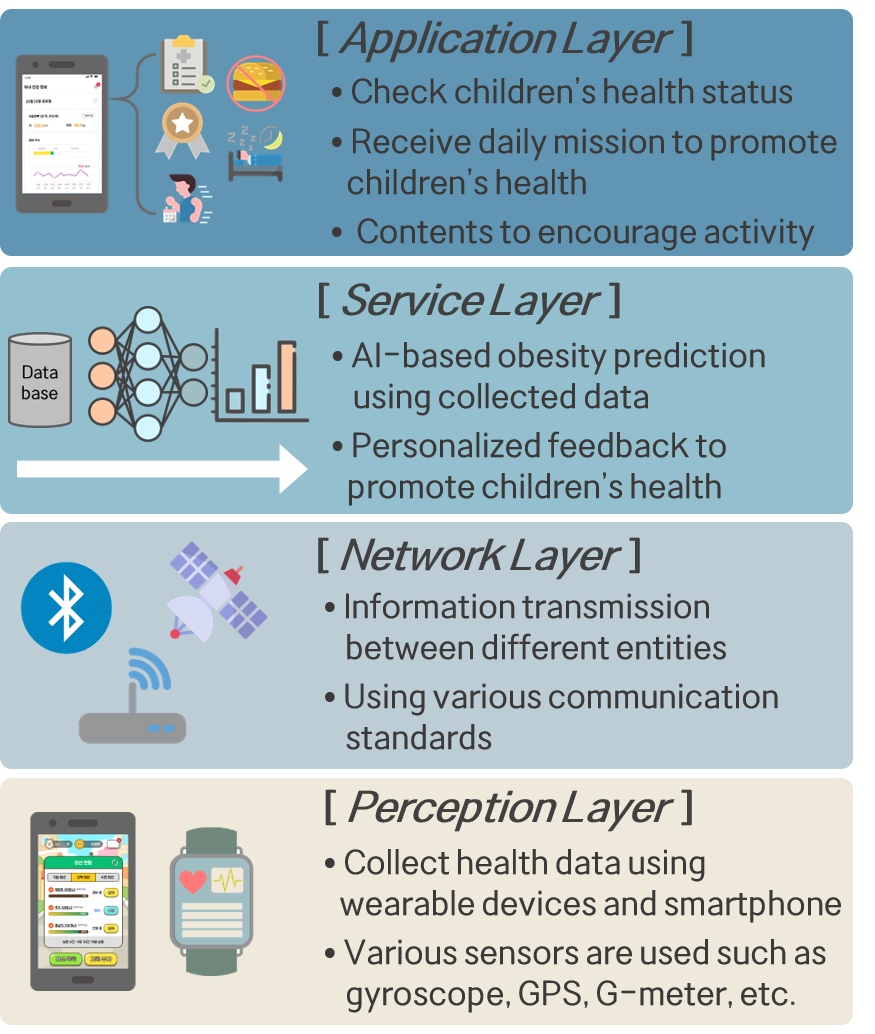}
\caption{\textcolor{black}{Architecture of WUDI with four layers}}
\label{fig_SOAArchitecture}
\end{figure}

The proposed self-adaptive system focused on healthcare services to prevent child obesity, thus IoT service-oriented architecture (SOA) \cite{islam2022internet} was applied. The architecture consists of four layers: 1) perception 2) network 3) service and 4) application layers. Fig. \ref{fig_SOAArchitecture} shows the layers of this architecture. 

The perception layer, also as known as the sensor layer, is responsible for collecting data, and WUDI collects lifelog using wearable devices and smartphones via a healthcare application. Various data were collected such as gyroscope, GPS, G-meter, heart rate, sleep time, step count, and physical information of children. 

The network layer (i.e., communication layer) is the core of IoT ecosystems. It provides connectivity using various network standards and supports information exchange between different objects \cite{lee2021survey, islam2022internet}. In WUDI, information exchange occurs between wearable devices, smartphones, and cloud servers.

In general, the service layer ensures sufficient service to assist the application layer \cite{lombardi2021internet}; therefore, in the service layer of WUDI, AI model training is performed, and the trained model is applied to predict child obesity using collected data from the perception layer. Personalized feedback (e.g., reward, daily mission) is generated in the service layer.

The application layer provides interfaces between users and applications; it complies with user requests for services \cite{islam2022internet}. The application layer provides health reports of children to their guardians, and children receive daily missions (i.e., daily workload) and rewards after exercise. 
\subsection{Self-adaptive Framework with Human-in-the-loop } \label{sec_selfAdaptiveFramework}

\begin{figure*}[!t]
\centering
\includegraphics[width=16cm]{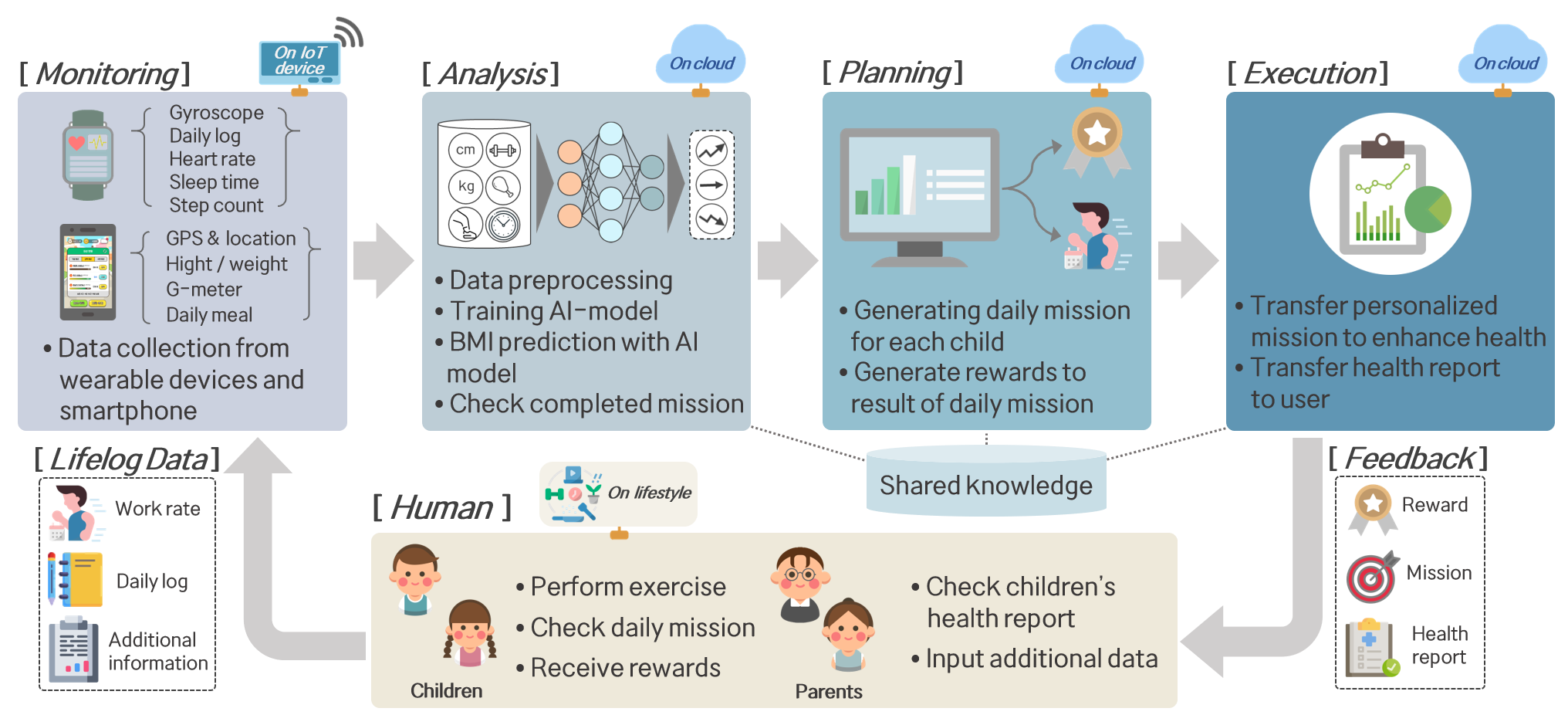}
\caption{Overview of the proposed framework with MAPE-H loop}
\label{fig_overview}
\end{figure*}

The proposed framework consists of the MAPE loop which is a prominent feedback loop used in the adaptation process in self-adaptive software and autonomous computing \cite{salehie2009self}. The MAPE loop consists of four processes: 1) monitoring, 2) analysis, 3) planning, and 4) execution. The human is located in the MAPE loop to interoperate with the healthcare system; thus, we named the loop MAPE-H loop. Fig. \ref{fig_overview} shows the overview of the proposed framework. 

In the MAPE-H loop, humans are classified as children and their guardians (e.g., parents and teachers), and the role of humans is similar to that of the sensor and actuator of IoT systems. In addition, the human can interact with a smartphone and smartwatch application in WUDI (Details about the smartphone application are described in Section \ref{sec_experiDataCollection}). As a sensor role, the children provided lifelogs through smartwatches and smartphones such as gyroscope, GPS, G-meter, heart rate, sleep time, and step count. As an actuator role, children performed exercises to accomplish their daily missions (i.e., recommend workload). Moreover, the children received rewards such as emoticons and special characters in the healthcare application if they continuously completed their daily missions. The guardians provided data such as height, weight, and daily diet as sensor roles, and received their children’s health reports. 

The monitoring process collected data from the smartwatch and smartphone applications. As described in the previous section, the data is classified as lifelog which is collected automatically such as gyroscope, heart rate, sleep time, step counts, GPS and G-meter. The data that could not be collected automatically were collected manually, such as physical information (e.g., height and weight) and daily meals and snacks. The collected data were transferred to the analysis process. 

The main purpose of the analysis was to analyze the collected data from the previous process. An AI-based method was used to predict child obesity; thus, general steps for AI were performed in the analysis such as data preprocessing and training the AI model. The trained AI model was used to predict child obesity. The details of this model are described in Section \ref{sec_ensembleModel}. In addition, the collected data and results of the AI model were saved in the shared database.

The planning process developed optimal strategies for each child to reduce obesity. Therefore, the strategy included generating daily missions (i.e., daily workload) to reduce weight and suitable rewards to encourage children. WUDI did not recommend reducing diet because it can potentially have negative nutrition effects on children. The generated strategies were transferred to the next process. The information generated in the planning process was managed in the shared database. 

The execution process delivered the strategies (i.e., daily missions and rewards) to the children. Health reports were delivered to the guardians, including body changes, mission completion, eaten calories, consumed calories, and sleep time. After the execution, the human received results as a part of MAPE-H. Then, the new data was generated and transferred to the monitoring process, and MAPE-H loop was continued.
\subsection{Obesity Prediction with Ensemble-based Model} \label{sec_ensembleModel}

\begin{figure*}[!t]
\centering
\includegraphics[width=15cm]{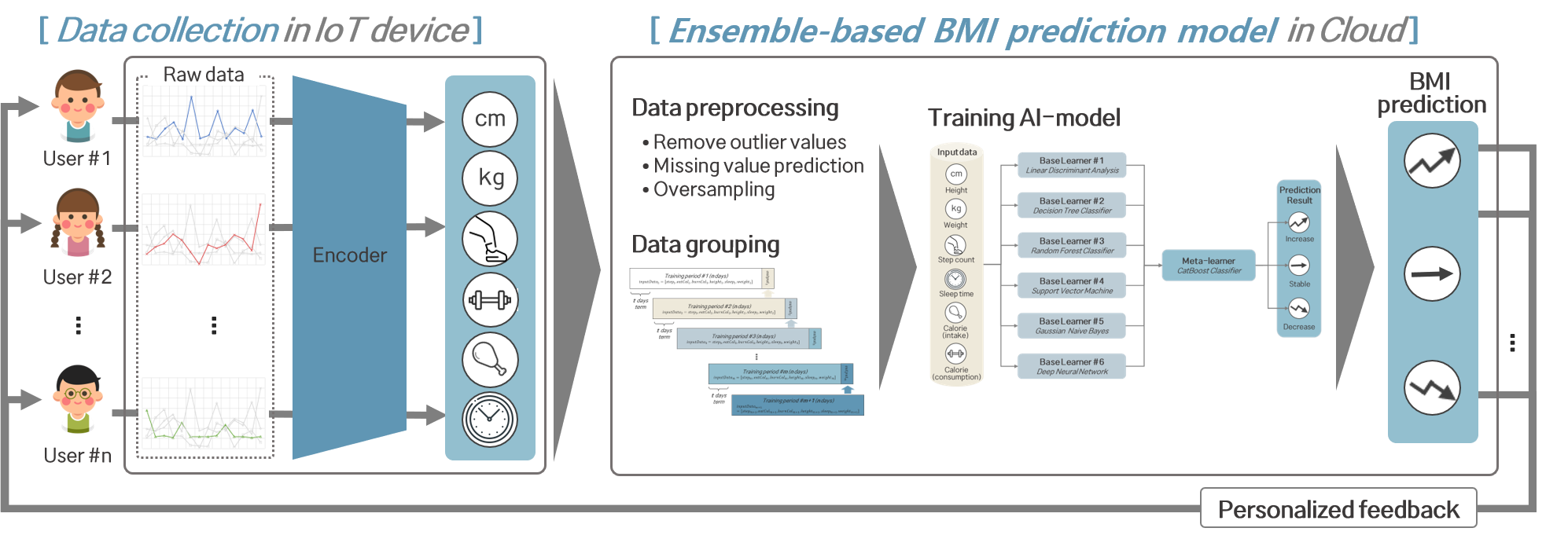}
\caption{Overview of ensemble-based machine learning to predict child obesity}
\label{fig_EnsembleOverview}
\end{figure*}

In this section, machine-learning processes to predict obesity in WUDI are described; the details described are between the machine-learning process and MAPE-H loop. Fig. \ref{fig_EnsembleOverview} depicts the processes, and the processes are divided into data collection and machine-learning phases.

\subsubsection{Data collection}
Machine learning requires a large amount of data to generate accurate models; the collection of this data is performed preemptively. The data collection is performed in the monitoring process and the human side of the MAPE-H loop. Raw data were collected in two ways: life logs using smart devices and directly entered data. The lifelog data were collected as raw data through smartphone and smartwatch applications such as gyroscope, GPS, G-meter, motion detection, and usage of smart devices. The other raw data, such as height, weight, and daily meals and snacks were direct input by the user. 

The collected raw data was encoded as the learning features to predict obesity, and the features are described as follows: 
\begin{itemize}
    \item \textit{Height} and \textit{weight} are the factors used to calculate BMI which is a metric of obesity; thus, it was used as the feature for machine learning \cite{huffman2010parenthood}.
    \item \textit{Calories intake} can affect obesity if the intake is excessive \cite{huffman2010parenthood}. 
    \item \textit{Exercise} and \textit{daily step count} can measure consumed calories \cite{dugan2008exercise, sothern2001exercise}. In WUDI, the consumed calories are extracted using the metabolic equivalent of task (MET) which is the objective measurement of expended energy from physical activity \cite{mendes2018metabolic, haskell2007physical}. The physical activities were classified into eight categories (i.e., walking, jumping jacks, jumping rope, hula hoop, starching, running, doing sit-ups, and discus throwing). 
    \item \textit{Sleep duration} is also associated with childhood obesity \cite{chen2008sleep, snell2007sleep}; thus, it was used in the features that predict BMI change.
\end{itemize}

The encoded features were transferred to the cloud server (i.e., MAPE-H analysis). However, the details of the encoding process from raw data to the learning features are out of the scope of this article.

\subsubsection{Data Preprocessing and Grouping}
The obesity prediction model was trained in the cloud, and the training process was related to the analysis process in the MAPE-H loop. The transferred data from the data collection phase were refined by removing outliers and missing values, and the refined data were oversampled to address data imbalance using oversampling techniques such as synthetic minority over-sampling technique (SMOTE) \cite{chawla2002smote}, K-means SMOTE \cite{douzas2018improving}, and random oversampling. 

\begin{figure*}[!t]
\centering
\includegraphics[width=14cm]{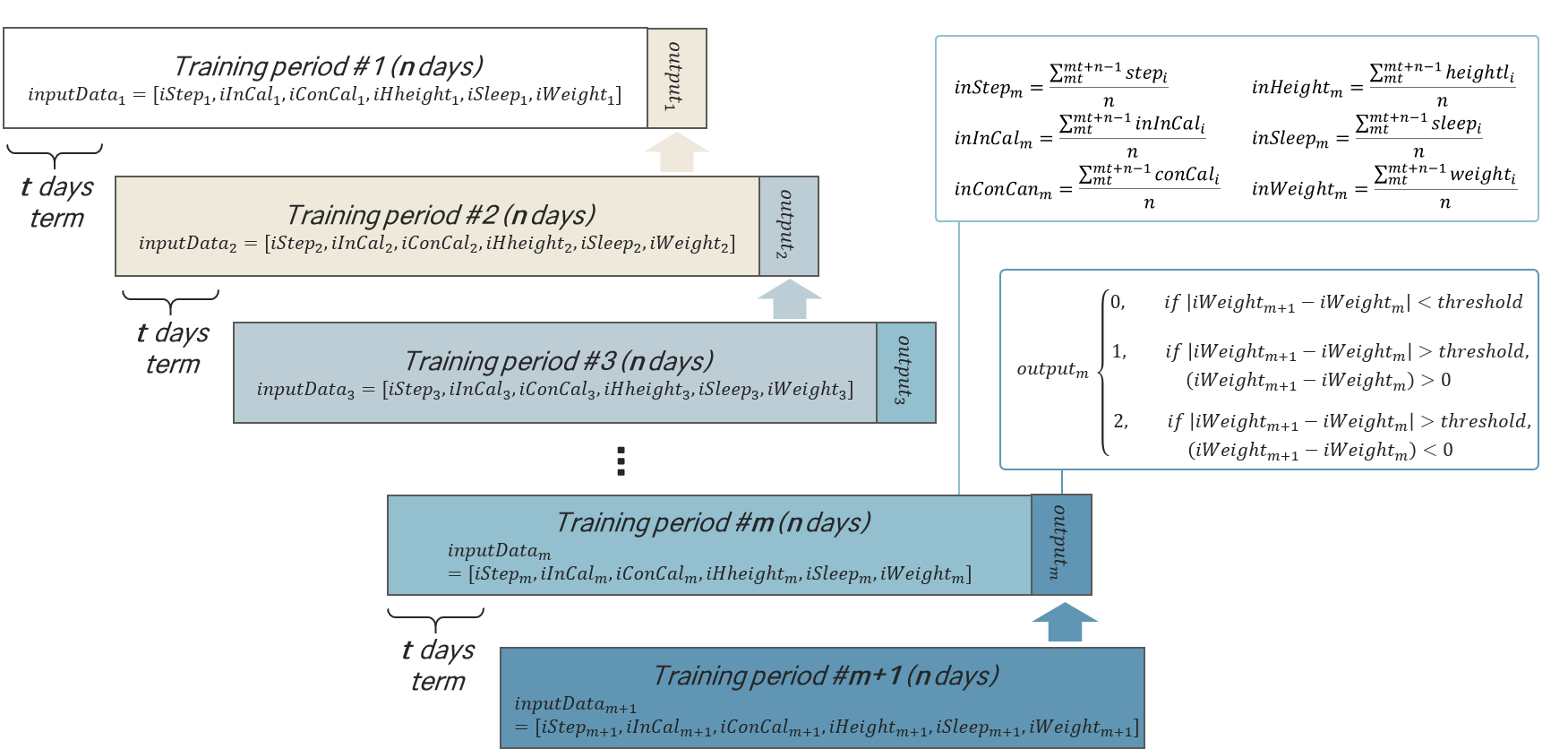}
\caption{A graphical depiction of the grouping of data}
\label{fig_groupping}
\end{figure*}

After preprocessing, the data were grouped to reflect the specific lifelog periods of each child. As BMI does not change instantly by exercising or diet management for a few days; thus, the grouping was required to reflect the change of specific duration. Fig. \ref{fig_groupping} shows the visual representation of the grouping.

The grouped data were used for training and testing to develop the model. Input data were the mean values of the features (i.e., height, weight, calorie intake, calorie consumption, sleep duration, and step count) in a specific period. If the period is set as $n$ days with t term days, $m$ th training data can be generated by following formula.

\begin{multline}
    inputData_{m}= [inStep_{m}, inInCal_{m}, inConCal_{m}, \\ inHeight_{m}, inSleep_{m}, inWeight_{m} ]
\end{multline}

Let the input features be collected daily and denoted as $feature_{days}$. The mean values of the features are calculated using the following equations. 

\begin{equation}
  \begin{cases}
    \ inStep_{m} = \cfrac{\sum_{mt}^{mt+n-1}step_{i}}{n} \\\\
    \ inInCal_{m} = \cfrac{\sum_{mt}^{mt+n-1}inInCal_{i}}{n} \\\\
    \ inConCal_{m} = \cfrac{\sum_{mt}^{mt+n-1}inConCal_{i}}{n} \\\\
    \ inHeight_{m} = \cfrac{\sum_{mt}^{mt+n-1}inHeight_{i}}{n} \\\\
    \ inSleep_{m} = \cfrac{\sum_{mt}^{mt+n-1}inSleep_{i}}{n} \\\\
    \ inWeight_{m} = \cfrac{\sum_{mt}^{mt+n-1}inWeight_{i}}{n} \\\\
  \end{cases}
\end{equation}

The output is the change in weight in the specific period; thus, the output is calculated using the difference with the weight value (i.e., $inWeight_{m}$) of the next input data (i.e., $inWeight_{m+1}$). The output is classified as maintain (0), increasing (1), and decreasing (2), and the output can be calculated using a threshold as follows. 

\begin{equation}
  output_{m}
  \begin{cases}
     \ 0, & if \left| weight_{m+1}-weight_{m} \right| < threshold \\
     \ 1, & if \left| weight_{m+1}-weight_{m} \right| > threshold, \\
         & \left( weight_{m+1}-weight_{m} \right) > threshold \\
     \ 2, & if \left| weight_{m+1}-weight_{m} \right| > threshold, \\
         & \left( weight_{m+1}-weight_{m} \right) < threshold \\
  \end{cases}
\end{equation}

Finally, the grouped input and output data are used to train and test the prediction model.

\subsubsection{Ensemble-based Learning Model for Obesity Prediction}

\begin{figure}[!t]
\centering
\includegraphics[width=9cm]{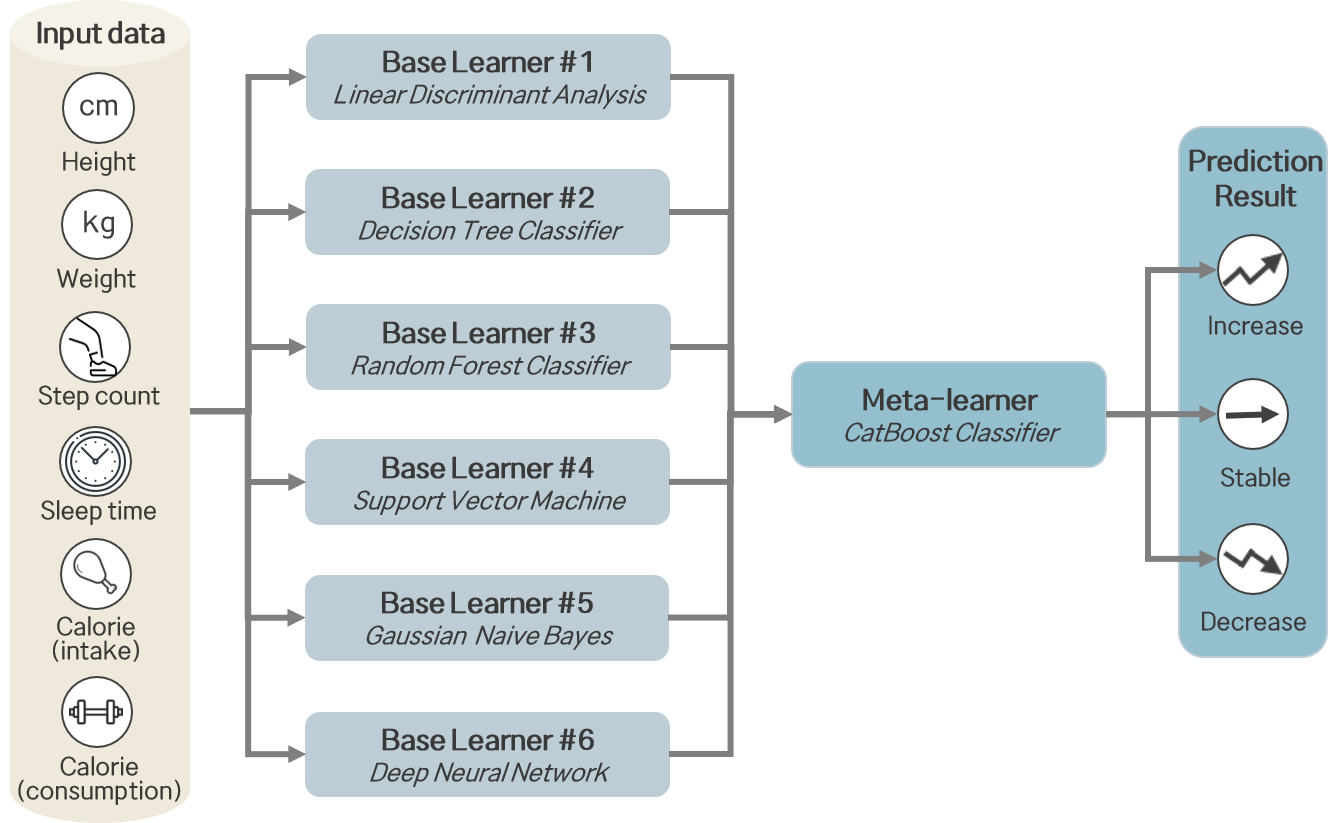}
\caption{Structure of the proposed deep ensemble learning model}
\label{fig_ensemble}
\end{figure}

We applied an ensemble-based learning method to predict obesity. Ensemble learning trains various machine-learning methods and combines prediction results to enhance prediction performance compared to that of the individual method \cite{mienye2022survey}. It has been applied in healthcare-related research such as kidney-stone detection \cite{kazemi2018novel}, diseases prediction \cite{fitriyani2019development}, breast cancer prediction \cite{gu2020case}, breast cancer classification \cite{ghiasi2021application}, and heart disease prediction \cite{mienye2020improved, gao2021improving, jothi2021enhanced}. Currently, deep-learning-based models perform better than traditional machine-learning methods; thus, the deep-learning model was combined with ensemble-based learning (i.e., ensemble deep learning) \cite{ganaie2022ensemble} to leverage the benefits of deep-learning ensembles. The structure of the proposed deep-ensemble-learning model is illustrated in Fig. \ref{fig_ensemble}. 

The proposed model applied a stacking approach, which leverages the abilities of multiple well-formed machine-learning algorithms to produce accurate classification results \cite{mienye2022survey}. Six base learners were selected in the proposed model: linear discriminant analysis \cite{mclachlan2005discriminant}, DT classifier \cite{priyanka2020decision}, random forest classifier \cite{breiman2001random}, SVM \cite{hearst1998support, noble2006support}, Gaussian naïve bayes \cite{hastie2009elements}, and deep neural network (DNN) \cite{russell2010artificial}. Let $T_{algorighm}[in, out]$   denote a training model that uses a specific algorithm with a set of inputs and outputs. The trained model for each base learner was generated by training on a set of k grouped data, as expressed below.

\begin{equation}
  \begin{cases}
    \ LDA = T_{LDA}[in_{1...k}, out_{1...k}] \\
    \ DTC = T_{DTC}[in_{1...k}, out_{1...k}] \\
    \ RFC = T_{RFC}[in_{1...k}, out_{1...k}] \\
    \ SVM = T_{SVM}[in_{1...k}, out_{1...k}] \\
    \ BNB = T_{BNB}[in_{1...k}, out_{1...k}] \\
    \ DNN = T_{DNN}[in_{1...k}, out_{1...k}] \\
  \end{cases}
\end{equation}

The predicted results of the base learners were used as input of a meta-learner that combined the prediction from the base learners. The CatBoost classifier \cite{prokhorenkova2018catboost} was employed as the meta-learner owing to its ability to effectively handle categorical features. The meta-learner was trained and tested with the prediction results of the base learners as input data, and the trained meta-leaner model determined the final prediction results. Therefore, the training of the proposed ensemble model is expressed as follows:

\begin{multline}
    WUDI_{ensemble}= \\ T_{CatBoost}[\{LDA(in_{1...k}), DTC(in_{1...k}), RFC(in_{1...k}), \\ SVM(in_{1...k}), BNB(in_{1...k}),  DNN(in_{1...k})\},out_{1...k}]
\end{multline}

The training and testing of the proposed ensemble model were also performed in the cloud as part of the analysis process in MAPE-H loop. The trained model was used to predict obesity of children using their n days (i.e., the criteria of data grouping) lifelog data for analysis. 

\section{empirical evaluation} \label{sec_empiricalEvaluation}

This section discussed a set of experiments for the empirical evaluation of the proposed framework. The experiments were conducted in Python 3.8.8 and PyTorch 1.8.1, and the test device configuration was Intel® Core™ i7-10700K (3.80 GHz and 8 cores), 32 GB memory, GeForce RTX 3070, and Windows 10. Experimental data were collected via implemented applications and the details are described in Section \ref{sec_experiDataCollection}. The results of the experiments on obesity prediction and effects of reward are described in Section \ref{sec_resultObesityPredict} and \ref{sec_resultRewards} respectively.

\subsection{Experimental Data Collection} \label{sec_experiDataCollection}

Data was collected using smartphone applications. The applications were implemented in Android and iOS, and the applications were launched in Google play\footnote{https://play.google.com/store/apps/details?id=com.injewelme.wydi} and App store\footnote{https://apps.apple.com/kr/app/id1548250829}. We selected a cohort of 362 Korean children aged between 104 and 152 months and provided smartwatches (i.e., Galaxy Fit2) \cite{GalaxyFit2} for them to obtain precise lifelog data. Daily lifelog data was collected over a period of three months, while body information (height and weight) was physically measured by experts at the beginning and end of the experimental period. The collected lifelog data were translated into input features (i.e., height, weight, calorie intake, consumed calorie, sleep duration, and step count) and grouped on a daily basis. Finally, the experimental dataset containing daily lifelog data of 316 children over three months was collected after data preprocessing. Note that the information was de-identified in the applications. Also, the overall experimental protocols and environments were reviewed and approved by the Institutional Review Board (IRB) of Chungbuk National University (CBNU-202308-HR-0196),

\subsection{Results of Obesity Prediction } \label{sec_resultObesityPredict}

We conducted data preprocessing and model configuration. SMOTE \cite{douzas2018improving}, was applied to adjust the distribution of the experimental data. The collected data were grouped as five days, and three-day terms were applied among groups (see Section \ref{sec_ensembleModel}) 

\begin{figure}[!t]
\centering
\includegraphics[width=6.5cm]{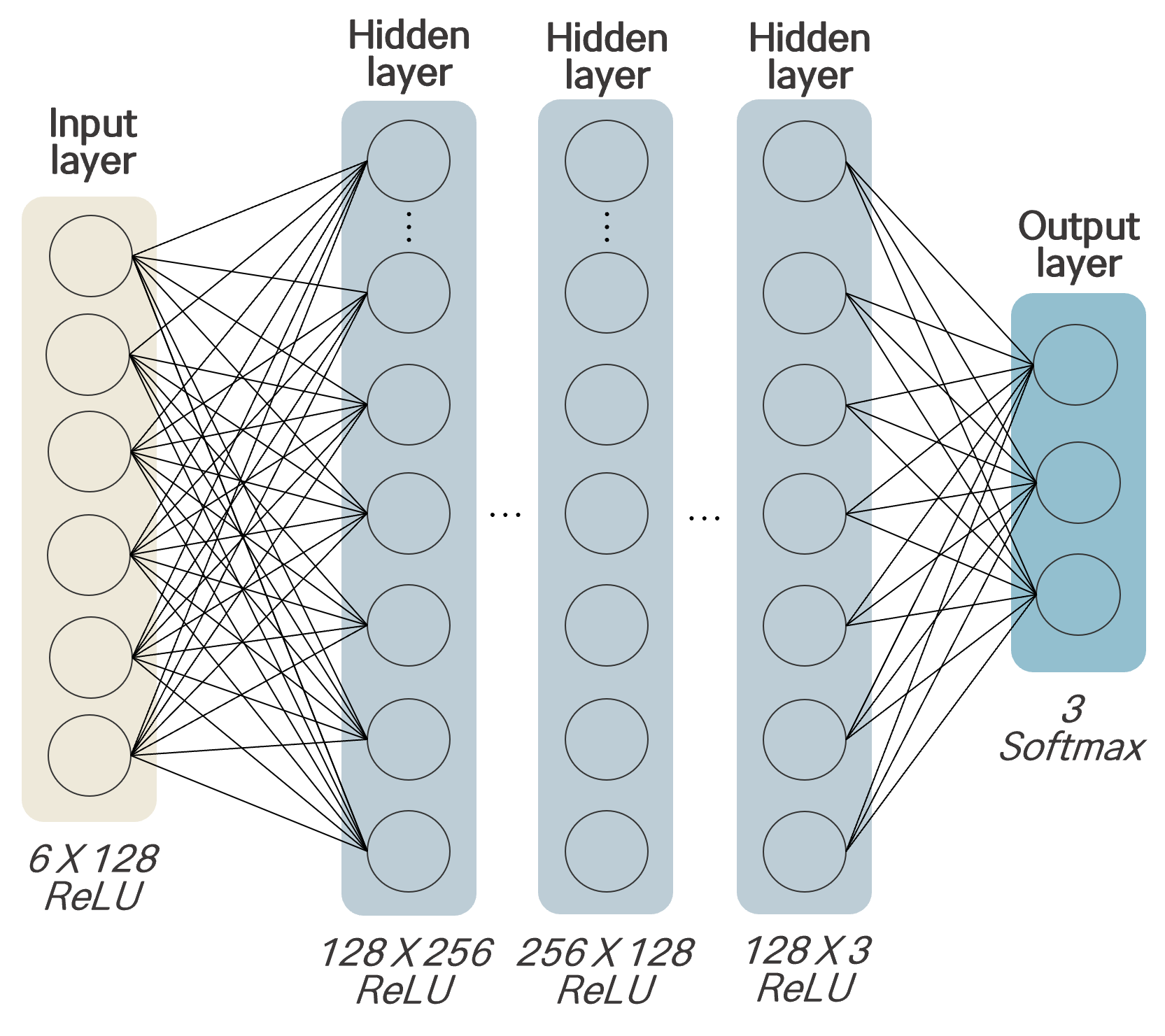}
\caption{DNN model for experiment}
\label{fig_DNN}
\end{figure}

In WUDI, the ensemble-based obesity prediction model is applied, and the model consists six base learners. Each base learner was optimized to perform the best prediction for experiments. Especially, the DNN-based learner was composed of a fully connected network with three hidden layers. ReLU [\cite{li2017convergence}, Adam \cite{zhang2018improved}, cross-entropy loss \cite{zhang2018generalized}, and softmax function \cite{russell2010artificial} were applied as activation function, loss function, and optimizer.  Fig. \ref{fig_DNN} shows the graphical representation of the DNN. 

\begin{figure*}[!t]
\centering
\includegraphics[width=12.8cm]{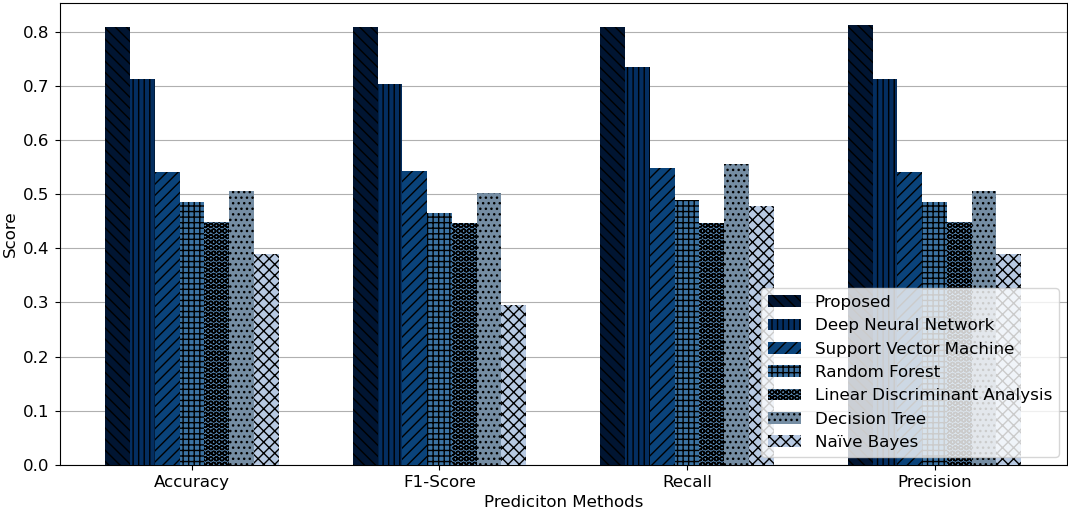}
\caption{Comparison results of obesity prediction}
\label{fig_expePredict}
\end{figure*}

To evaluate the efficiency of the proposed ensemble-based obesity prediction model, k-fold cross-validation \cite{hastie2009elements} was applied to evaluate generalizability. The data was divided into 10 equal subsets, and 80\% of the data was used for training the model. The base learners were compared with the ensemble model in prediction performance. Accuracy, precision, recall, and F1-score were applied as performance metrics \cite{russell2010artificial, hastie2009elements}. The prediction results for the k-fold cross-validation are presented in Table \ref{tbl_compareResearch}. Fig. \ref{fig_expePredict} shows the overall results for each measurement.

\begin{table*}
\scriptsize
\caption{Comparison results of prediction performance}
\label{tbl_compareResearch}
\centering
\begin{tabular}{|rlllllllllll|}
\hline
\multicolumn{12}{|c|}{\textbf{Accuracy}}\\ \hline
\multicolumn{1}{|l|}{}                             & \multicolumn{1}{l|}{1-fold}           & \multicolumn{1}{l|}{2-fold}           & \multicolumn{1}{l|}{3-fold}           & \multicolumn{1}{l|}{4-fold}           & \multicolumn{1}{l|}{5-fold}           & \multicolumn{1}{l|}{6-fold}           & \multicolumn{1}{l|}{7-fold}           & \multicolumn{1}{l|}{8-fold}           & \multicolumn{1}{l|}{9-fold}           & \multicolumn{1}{l|}{10-fold}          & Average          \\ \hline
\multicolumn{1}{|r|}{\textbf{Proposed}}            & \multicolumn{1}{l|}{\textbf{0.66254}} & \multicolumn{1}{l|}{\textbf{0.78602}} & \multicolumn{1}{l|}{\textbf{0.82332}} & \multicolumn{1}{l|}{\textbf{0.84727}} & \multicolumn{1}{l|}{\textbf{0.79859}} & \multicolumn{1}{l|}{\textbf{0.85552}} & \multicolumn{1}{l|}{\textbf{0.82548}} & \multicolumn{1}{l|}{\textbf{0.81268}} & \multicolumn{1}{l|}{\textbf{0.82388}} & \multicolumn{1}{l|}{\textbf{0.84567}} & \textbf{0.80810} \\ \hline
\multicolumn{1}{|r|}{Deep Neural Network}          & \multicolumn{1}{l|}{0.49490}          & \multicolumn{1}{l|}{0.69356}          & \multicolumn{1}{l|}{0.73793}          & \multicolumn{1}{l|}{0.76345}          & \multicolumn{1}{l|}{0.68865}          & \multicolumn{1}{l|}{0.77091}          & \multicolumn{1}{l|}{0.72261}          & \multicolumn{1}{l|}{0.73218}          & \multicolumn{1}{l|}{0.73827}          & \multicolumn{1}{l|}{0.77459}          & 0.71170          \\ \hline
\multicolumn{1}{|r|}{Support Vector Machine}       & \multicolumn{1}{l|}{0.57833}          & \multicolumn{1}{l|}{0.51983}          & \multicolumn{1}{l|}{0.53965}          & \multicolumn{1}{l|}{0.56675}          & \multicolumn{1}{l|}{0.51276}          & \multicolumn{1}{l|}{0.58049}          & \multicolumn{1}{l|}{0.50510}          & \multicolumn{1}{l|}{0.51659}          & \multicolumn{1}{l|}{0.53878}          & \multicolumn{1}{l|}{0.55174}          & 0.54100          \\ \hline
\multicolumn{1}{|r|}{Random Forest}                & \multicolumn{1}{l|}{0.51139}          & \multicolumn{1}{l|}{0.48685}          & \multicolumn{1}{l|}{0.48842}          & \multicolumn{1}{l|}{0.48312}          & \multicolumn{1}{l|}{0.47487}          & \multicolumn{1}{l|}{0.48822}          & \multicolumn{1}{l|}{0.44680}          & \multicolumn{1}{l|}{0.47771}          & \multicolumn{1}{l|}{0.50599}          & \multicolumn{1}{l|}{0.49578}          & 0.48591          \\ \hline
\multicolumn{1}{|r|}{Linear Discriminant Analysis} & \multicolumn{1}{l|}{0.47527}          & \multicolumn{1}{l|}{0.45426}          & \multicolumn{1}{l|}{0.45858}          & \multicolumn{1}{l|}{0.42108}          & \multicolumn{1}{l|}{0.43659}          & \multicolumn{1}{l|}{0.43738}          & \multicolumn{1}{l|}{0.41676}          & \multicolumn{1}{l|}{0.44551}          & \multicolumn{1}{l|}{0.47006}          & \multicolumn{1}{l|}{0.46142}          & 0.44769          \\ \hline
\multicolumn{1}{|r|}{Decision Tree}                & \multicolumn{1}{l|}{0.52022}          & \multicolumn{1}{l|}{0.51197}          & \multicolumn{1}{l|}{0.51394}          & \multicolumn{1}{l|}{0.56007}          & \multicolumn{1}{l|}{0.45190}          & \multicolumn{1}{l|}{0.55752}          & \multicolumn{1}{l|}{0.43856}          & \multicolumn{1}{l|}{0.51581}          & \multicolumn{1}{l|}{0.50560}          & \multicolumn{1}{l|}{0.48773}          & 0.50633          \\ \hline
\multicolumn{1}{|r|}{Naïve Bayes}                  & \multicolumn{1}{l|}{0.38732}          & \multicolumn{1}{l|}{0.41186}          & \multicolumn{1}{l|}{0.40891}          & \multicolumn{1}{l|}{0.40185}          & \multicolumn{1}{l|}{0.35925}          & \multicolumn{1}{l|}{0.42246}          & \multicolumn{1}{l|}{0.34197}          & \multicolumn{1}{l|}{0.39171}          & \multicolumn{1}{l|}{0.39446}          & \multicolumn{1}{l|}{0.37699}          & 0.38968          \\ \hline
\multicolumn{12}{|c|}{\textbf{F1-Score}}                                                                                                                                                                                                                                                                                                                                                                                                                                              \\ \hline
\multicolumn{1}{|l|}{}                             & \multicolumn{1}{l|}{1-fold}           & \multicolumn{1}{l|}{2-fold}           & \multicolumn{1}{l|}{3-fold}           & \multicolumn{1}{l|}{4-fold}           & \multicolumn{1}{l|}{5-fold}           & \multicolumn{1}{l|}{6-fold}           & \multicolumn{1}{l|}{7-fold}           & \multicolumn{1}{l|}{8-fold}           & \multicolumn{1}{l|}{9-fold}           & \multicolumn{1}{l|}{10-fold}          & Average          \\ \hline
\multicolumn{1}{|r|}{\textbf{Proposed}}            & \multicolumn{1}{l|}{\textbf{0.66414}} & \multicolumn{1}{l|}{\textbf{0.78497}} & \multicolumn{1}{l|}{\textbf{0.82315}} & \multicolumn{1}{l|}{\textbf{0.84684}} & \multicolumn{1}{l|}{\textbf{0.79782}} & \multicolumn{1}{l|}{\textbf{0.85404}} & \multicolumn{1}{l|}{\textbf{0.82468}} & \multicolumn{1}{l|}{\textbf{0.81199}} & \multicolumn{1}{l|}{\textbf{0.82372}} & \multicolumn{1}{l|}{\textbf{0.84583}} & \textbf{0.80772} \\ \hline
\multicolumn{1}{|r|}{Deep Neural Network}          & \multicolumn{1}{l|}{0.42731}          & \multicolumn{1}{l|}{0.68886}          & \multicolumn{1}{l|}{0.73675}          & \multicolumn{1}{l|}{0.76281}          & \multicolumn{1}{l|}{0.68330}          & \multicolumn{1}{l|}{0.76926}          & \multicolumn{1}{l|}{0.72132}          & \multicolumn{1}{l|}{0.73067}          & \multicolumn{1}{l|}{0.73733}          & \multicolumn{1}{l|}{0.77372}          & 0.70313          \\ \hline
\multicolumn{1}{|r|}{Support Vector Machine}       & \multicolumn{1}{l|}{0.57665}          & \multicolumn{1}{l|}{0.52133}          & \multicolumn{1}{l|}{0.54105}          & \multicolumn{1}{l|}{0.56758}          & \multicolumn{1}{l|}{0.51413}          & \multicolumn{1}{l|}{0.58095}          & \multicolumn{1}{l|}{0.50410}          & \multicolumn{1}{l|}{0.51773}          & \multicolumn{1}{l|}{0.54073}          & \multicolumn{1}{l|}{0.55323}          & 0.54175          \\ \hline
\multicolumn{1}{|r|}{Random Forest}                & \multicolumn{1}{l|}{0.49186}          & \multicolumn{1}{l|}{0.46681}          & \multicolumn{1}{l|}{0.47079}          & \multicolumn{1}{l|}{0.46192}          & \multicolumn{1}{l|}{0.45479}          & \multicolumn{1}{l|}{0.46699}          & \multicolumn{1}{l|}{0.42110}          & \multicolumn{1}{l|}{0.46126}          & \multicolumn{1}{l|}{0.48625}          & \multicolumn{1}{l|}{0.47547}          & 0.46572          \\ \hline
\multicolumn{1}{|r|}{Linear Discriminant Analysis} & \multicolumn{1}{l|}{0.47034}          & \multicolumn{1}{l|}{0.45311}          & \multicolumn{1}{l|}{0.45498}          & \multicolumn{1}{l|}{0.42140}          & \multicolumn{1}{l|}{0.43638}          & \multicolumn{1}{l|}{0.43608}          & \multicolumn{1}{l|}{0.41669}          & \multicolumn{1}{l|}{0.44429}          & \multicolumn{1}{l|}{0.46598}          & \multicolumn{1}{l|}{0.45915}          & 0.44584          \\ \hline
\multicolumn{1}{|r|}{Decision Tree}                & \multicolumn{1}{l|}{0.52743}          & \multicolumn{1}{l|}{0.51038}          & \multicolumn{1}{l|}{0.51475}          & \multicolumn{1}{l|}{0.56484}          & \multicolumn{1}{l|}{0.44010}          & \multicolumn{1}{l|}{0.56282}          & \multicolumn{1}{l|}{0.40424}          & \multicolumn{1}{l|}{0.50527}          & \multicolumn{1}{l|}{0.49502}          & \multicolumn{1}{l|}{0.49239}          & 0.50173          \\ \hline
\multicolumn{1}{|r|}{Naïve Bayes}                  & \multicolumn{1}{l|}{0.27648}          & \multicolumn{1}{l|}{0.31848}          & \multicolumn{1}{l|}{0.31276}          & \multicolumn{1}{l|}{0.31095}          & \multicolumn{1}{l|}{0.26017}          & \multicolumn{1}{l|}{0.32659}          & \multicolumn{1}{l|}{0.25084}          & \multicolumn{1}{l|}{0.30722}          & \multicolumn{1}{l|}{0.30366}          & \multicolumn{1}{l|}{0.28807}          & 0.29552          \\ \hline
\multicolumn{12}{|c|}{\textbf{Precision}}                                                                                                                                                                                                                                                                                                                                                                                                                                             \\ \hline
\multicolumn{1}{|l|}{}                             & \multicolumn{1}{l|}{1-fold}           & \multicolumn{1}{l|}{2-fold}           & \multicolumn{1}{l|}{3-fold}           & \multicolumn{1}{l|}{4-fold}           & \multicolumn{1}{l|}{5-fold}           & \multicolumn{1}{l|}{6-fold}           & \multicolumn{1}{l|}{7-fold}           & \multicolumn{1}{l|}{8-fold}           & \multicolumn{1}{l|}{9-fold}           & \multicolumn{1}{l|}{10-fold}          & Average          \\ \hline
\multicolumn{1}{|r|}{\textbf{Proposed}}            & \multicolumn{1}{l|}{\textbf{0.66700}} & \multicolumn{1}{l|}{\textbf{0.78834}} & \multicolumn{1}{l|}{\textbf{0.82902}} & \multicolumn{1}{l|}{\textbf{0.84911}} & \multicolumn{1}{l|}{\textbf{0.80879}} & \multicolumn{1}{l|}{\textbf{0.85610}} & \multicolumn{1}{l|}{\textbf{0.82966}} & \multicolumn{1}{l|}{\textbf{0.81555}} & \multicolumn{1}{l|}{\textbf{0.82565}} & \multicolumn{1}{l|}{\textbf{0.85126}} & \textbf{0.81205} \\ \hline
\multicolumn{1}{|r|}{Deep Neural Network}          & \multicolumn{1}{l|}{0.60000}          & \multicolumn{1}{l|}{0.71431}          & \multicolumn{1}{l|}{0.74828}          & \multicolumn{1}{l|}{0.77576}          & \multicolumn{1}{l|}{0.71449}          & \multicolumn{1}{l|}{0.77796}          & \multicolumn{1}{l|}{0.73715}          & \multicolumn{1}{l|}{0.74837}          & \multicolumn{1}{l|}{0.75134}          & \multicolumn{1}{l|}{0.78689}          & 0.73545          \\ \hline
\multicolumn{1}{|r|}{Support Vector Machine}       & \multicolumn{1}{l|}{0.57802}          & \multicolumn{1}{l|}{0.52634}          & \multicolumn{1}{l|}{0.54557}          & \multicolumn{1}{l|}{0.57006}          & \multicolumn{1}{l|}{0.52400}          & \multicolumn{1}{l|}{0.58594}          & \multicolumn{1}{l|}{0.51273}          & \multicolumn{1}{l|}{0.52699}          & \multicolumn{1}{l|}{0.54718}          & \multicolumn{1}{l|}{0.55697}          & 0.54738          \\ \hline
\multicolumn{1}{|r|}{Random Forest}                & \multicolumn{1}{l|}{0.51352}          & \multicolumn{1}{l|}{0.49454}          & \multicolumn{1}{l|}{0.48184}          & \multicolumn{1}{l|}{0.48071}          & \multicolumn{1}{l|}{0.47558}          & \multicolumn{1}{l|}{0.49738}          & \multicolumn{1}{l|}{0.45982}          & \multicolumn{1}{l|}{0.47969}          & \multicolumn{1}{l|}{0.50539}          & \multicolumn{1}{l|}{0.49613}          & 0.48846          \\ \hline
\multicolumn{1}{|r|}{Linear Discriminant Analysis} & \multicolumn{1}{l|}{0.47232}          & \multicolumn{1}{l|}{0.45335}          & \multicolumn{1}{l|}{0.45420}          & \multicolumn{1}{l|}{0.42306}          & \multicolumn{1}{l|}{0.43680}          & \multicolumn{1}{l|}{0.43814}          & \multicolumn{1}{l|}{0.41704}          & \multicolumn{1}{l|}{0.44465}          & \multicolumn{1}{l|}{0.46689}          & \multicolumn{1}{l|}{0.45804}          & 0.44645          \\ \hline
\multicolumn{1}{|r|}{Decision Tree}                & \multicolumn{1}{l|}{0.56912}          & \multicolumn{1}{l|}{0.58395}          & \multicolumn{1}{l|}{0.57692}          & \multicolumn{1}{l|}{0.60511}          & \multicolumn{1}{l|}{0.48967}          & \multicolumn{1}{l|}{0.57840}          & \multicolumn{1}{l|}{0.46431}          & \multicolumn{1}{l|}{0.60076}          & \multicolumn{1}{l|}{0.56960}          & \multicolumn{1}{l|}{0.51973}          & 0.55576          \\ \hline
\multicolumn{1}{|r|}{Naïve Bayes}                  & \multicolumn{1}{l|}{0.49982}          & \multicolumn{1}{l|}{0.44706}          & \multicolumn{1}{l|}{0.46318}          & \multicolumn{1}{l|}{0.56467}          & \multicolumn{1}{l|}{0.41201}          & \multicolumn{1}{l|}{0.45211}          & \multicolumn{1}{l|}{0.43800}          & \multicolumn{1}{l|}{0.53188}          & \multicolumn{1}{l|}{0.49298}          & \multicolumn{1}{l|}{0.48305}          & 0.47848          \\ \hline
\multicolumn{12}{|c|}{\textbf{Recall}}                                                                                                                                                                                                                                                                                                                                                                                                                                                \\ \hline
\multicolumn{1}{|l|}{}                             & \multicolumn{1}{l|}{1-fold}           & \multicolumn{1}{l|}{2-fold}           & \multicolumn{1}{l|}{3-fold}           & \multicolumn{1}{l|}{4-fold}           & \multicolumn{1}{l|}{5-fold}           & \multicolumn{1}{l|}{6-fold}           & \multicolumn{1}{l|}{7-fold}           & \multicolumn{1}{l|}{8-fold}           & \multicolumn{1}{l|}{9-fold}           & \multicolumn{1}{l|}{10-fold}          & Average          \\ \hline
\multicolumn{1}{|r|}{\textbf{Proposed}}            & \multicolumn{1}{l|}{\textbf{0.66254}} & \multicolumn{1}{l|}{\textbf{0.78602}} & \multicolumn{1}{l|}{\textbf{0.82332}} & \multicolumn{1}{l|}{\textbf{0.84727}} & \multicolumn{1}{l|}{\textbf{0.79859}} & \multicolumn{1}{l|}{\textbf{0.85552}} & \multicolumn{1}{l|}{\textbf{0.82548}} & \multicolumn{1}{l|}{\textbf{0.81268}} & \multicolumn{1}{l|}{\textbf{0.82388}} & \multicolumn{1}{l|}{\textbf{0.84567}} & \textbf{0.80810} \\ \hline
\multicolumn{1}{|r|}{Deep Neural Network}          & \multicolumn{1}{l|}{0.49490}          & \multicolumn{1}{l|}{0.69356}          & \multicolumn{1}{l|}{0.73793}          & \multicolumn{1}{l|}{0.76345}          & \multicolumn{1}{l|}{0.68865}          & \multicolumn{1}{l|}{0.77091}          & \multicolumn{1}{l|}{0.72262}          & \multicolumn{1}{l|}{0.73218}          & \multicolumn{1}{l|}{0.73827}          & \multicolumn{1}{l|}{0.77459}          & 0.71170          \\ \hline
\multicolumn{1}{|r|}{Support Vector Machine}       & \multicolumn{1}{l|}{0.57833}          & \multicolumn{1}{l|}{0.51983}          & \multicolumn{1}{l|}{0.53965}          & \multicolumn{1}{l|}{0.56675}          & \multicolumn{1}{l|}{0.51276}          & \multicolumn{1}{l|}{0.58049}          & \multicolumn{1}{l|}{0.50510}          & \multicolumn{1}{l|}{0.51659}          & \multicolumn{1}{l|}{0.53878}          & \multicolumn{1}{l|}{0.55174}          & 0.54100          \\ \hline
\multicolumn{1}{|r|}{Random Forest}                & \multicolumn{1}{l|}{0.51139}          & \multicolumn{1}{l|}{0.48685}          & \multicolumn{1}{l|}{0.48842}          & \multicolumn{1}{l|}{0.48312}          & \multicolumn{1}{l|}{0.47487}          & \multicolumn{1}{l|}{0.48822}          & \multicolumn{1}{l|}{0.44680}          & \multicolumn{1}{l|}{0.47772}          & \multicolumn{1}{l|}{0.50599}          & \multicolumn{1}{l|}{0.49578}          & 0.48591          \\ \hline
\multicolumn{1}{|r|}{Linear Discriminant Analysis} & \multicolumn{1}{l|}{0.47527}          & \multicolumn{1}{l|}{0.45426}          & \multicolumn{1}{l|}{0.45858}          & \multicolumn{1}{l|}{0.42108}          & \multicolumn{1}{l|}{0.43659}          & \multicolumn{1}{l|}{0.43738}          & \multicolumn{1}{l|}{0.41677}          & \multicolumn{1}{l|}{0.44551}          & \multicolumn{1}{l|}{0.47006}          & \multicolumn{1}{l|}{0.46142}          & 0.44769          \\ \hline
\multicolumn{1}{|r|}{Decision Tree}                & \multicolumn{1}{l|}{0.52022}          & \multicolumn{1}{l|}{0.51198}          & \multicolumn{1}{l|}{0.51394}          & \multicolumn{1}{l|}{0.56007}          & \multicolumn{1}{l|}{0.45190}          & \multicolumn{1}{l|}{0.55752}          & \multicolumn{1}{l|}{0.43856}          & \multicolumn{1}{l|}{0.51581}          & \multicolumn{1}{l|}{0.50560}          & \multicolumn{1}{l|}{0.48773}          & 0.50633          \\ \hline
\multicolumn{1}{|r|}{Naïve Bayes}                  & \multicolumn{1}{l|}{0.38732}          & \multicolumn{1}{l|}{0.41186}          & \multicolumn{1}{l|}{0.40891}          & \multicolumn{1}{l|}{0.40185}          & \multicolumn{1}{l|}{0.35925}          & \multicolumn{1}{l|}{0.42246}          & \multicolumn{1}{l|}{0.34197}          & \multicolumn{1}{l|}{0.39171}          & \multicolumn{1}{l|}{0.39446}          & \multicolumn{1}{l|}{0.37699}          & 0.38968          \\ \hline
\end{tabular}
\end{table*}

The experimental results demonstrate that the proposed ensemble-based obesity prediction model outperformed other models in all metrics, with an accuracy of 80.8\% in k-fold cross-validation. The results show that the proposed model is reliable and can be used to predict obesity in children with reasonable performance.

\subsection{Effect of Reward in MAPE-H} \label{sec_resultRewards}

The WUDI framework comprises the MAPE-H loop, which necessitates human intervention to achieve the objectives of the framework. Human intervention contributes to the success of the framework; thus, rewards were required to encourage active participation. 

To evaluate the relationship between reward and human intervention, the 316 children were grouped as follows: those that received full rewards (26), basic rewards (79), and partial rewards (211). The group that received full rewards was provided with a higher level of incentives than the other groups. The group with basic rewards received more incentives than the partial rewards group but less than the full rewards group. The partial rewards group received the lowest level of incentives. The rewards were delivered when the children complete daily missions as described in the previous Section (See Section \ref{sec_selfAdaptiveFramework}). 

The principal objective of the WUDI framework was to enhance the health conditions of children by preventing obesity. Therefore, changes in BMI were used as an efficiency metric for evaluating the effectiveness of the framework. The change in BMI was measured by the difference in BMI throughout the experiment, and the efficiency was measured using n children as follows: 

\begin{multline}
    RewardEfficiency = \\ \cfrac{\sum_{1}^{n} BMI_{afterExperiment}(i) - BMI_{initial}(i)}{n}  
\end{multline}

\begin{figure*}[!t]
\centering
\includegraphics[width=12.8cm]{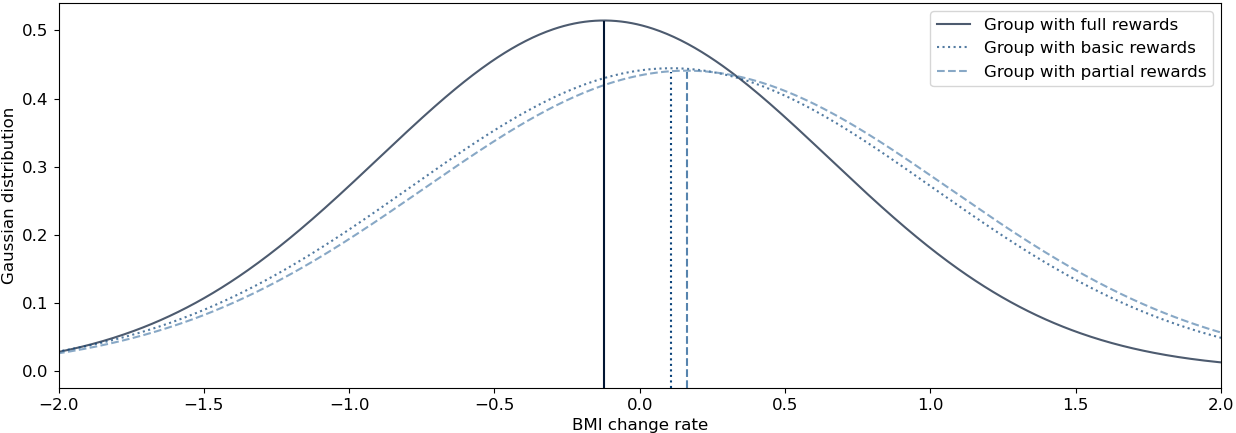}
\caption{Gaussian distribution of different reward groups}
\label{fig_expeReward}
\end{figure*}

Fig. \ref{fig_expeReward} shows the Gaussian distribution of reward efficiency of each reward group, and Table \ref{tbl_compareReward} shows the results of mean and standard deviation. 

\begin{table}[]
\caption{Result of efficiency of different reward groups}
\label{tbl_compareReward}
\begin{tabular}{|r|c|c|c|}
\hline
\multicolumn{1}{|l|}{} & Full rewards & Basic rewards & Partial rewards \\ \hline
Average                & -0.12333     & 0.10832       & 0.16151         \\ \hline
Standard deviation     & 0.77512      & 0.89753       & 0.90479         \\ \hline
\end{tabular}
\end{table}

As shown in the results, on average, the children who were fully rewarded experienced a reduction in their BMI. In contrast, the other groups increased BMI, on average, but the group that received more rewards had a slightly lower BMI increase. However, reducing BMI in growth periods is challenging because the BMI of Korean children steadily is increased with growth \cite{kim20182017}. Moreover, based on Korea National Health and Nutrition Examination Survey \cite{knhanes}, the means of BMI change of Korean children with the same constraints as the experimental dataset (i.e., children aged between 104 and 152 months, and three months BMI changes) were 0.17 (boys) and 0.15 (girls).

The results demonstrate that providing appropriate rewards can motivate children to reduce their weight by performing physical activity (i.e., daily mission) even without being centered on diet; thus, providing appropriate rewards can be applied to prevent childhood obesity in WUDI. The reward can serve as a motivator for human intervention and ultimately enhance the efficient operation of the MAPE-H loop.
\section{Discussion} \label{sec_discussion}

We proposed a self-adaptive healthcare framework to prevent childhood obesity, and the ensemble-based model was applied to predict obesity. Also, the rewards concept was applied to encourage human intervention in the MAPE-H loop. The proposed framework yielded excellent experimental results. However, the study has several limitations to be solved before moving forward. In this section, it is presented a discussion on the limitations and future research to overcome the limitations. 

The proposed childhood obesity model trained grouped data without considering the characteristics of each child. However, each student has different characteristics such as athletic ability and skeletal muscle mass. These individual characteristics can impact calorie consumption, and each child may consume different calories while performing the same exercises. Therefore, obesity prediction and reward provision have to consider these characteristics. To overcome the limitation, federated learning \cite{konevcny2016federated} may be applied. Federated learning trains a global model using aggregated individual models, and each individual trains their model using the global model and individual data; thus, the individual model can reflect overall and individual features of training. Therefore, if federated learning is applied to WUDI, individual models are generated for each child or group with similar characteristics. This is followed by the generation of a global model that aims to reflect the generality of children.

To predict childhood obesity, six factors were applied as features of the proposed obesity prediction model, and the experimental results showed reasonable efficiency of WUDI. Although the features significantly affect childhood obesity, the model must be considered to enhance the prediction results of a more precise obesity prediction. Various factors affect childhood obesity such as sex, race, eating habits, annual family income, education level of parent, family characteristics, community-related factors, and cultural environments \cite{sherburne2006review, huffman2010parenthood}. Future research should investigate the development of an obesity prediction model that considers a wider range of factors beyond the features used in this study.

Moreover, the proposed framework applied the MAPE-H loop to integrate humans as an element of the self-adaptive mechanism, and rewards were used to incentivize human involvement in weight management. Experimental results showed that rewards encourage willing participation. However, from the perspective of a service provider, excessive rewards can potentially interfere with the operation of WUDI services. Therefore, an accurate and precise rewards policy is necessary to strike a balance between incentivizing human involvement and ensuring profitable service operations. To address this limitation, future research can investigate reward recommendation techniques such as group \cite{seo2018enhanced} and personalized \cite{seo2017personalized} recommendations. Developing an accurate and effective reward policy can lead to a balance between encouraging human involvement and optimizing service operational profits.
\section {Conclusion} \label{sec_conclusion}
In this article, a self-adaptive healthcare framework to prevent childhood obesity using lifelog data was proposed and named WUDI. The framework applied SOA-based IoT architecture with four layers: application, service, network, and perception layers. The WUDI framework was designed to incorporate the self-adaptive concept and involve humans in a loop system; thus, WUDI integrated the human involvement and MAPE loop (i.e., MAPE-H loop). In the MAPE-H loop, the role of humans is similar to sensors and actuators of general IoT systems, and rewards concepts were applied to encourage human involvement. In WUDI, an ensemble-based learning model was proposed to predict childhood obesity, and six features were identified as features (i.e., height, weight, calorie intake, calorie consumed, daily step count, and sleep time) that can be extracted by sensors from smartphones and smartwatches.

For the experiment, lifelog data of children were collected using a smartphone application that provides real services. The experimental results highlighted the superiority of the proposed prediction model over other machine-learning models, were reasonable, and can be applied in healthcare service. Moreover, the experimental data classified three groups based on reward acquisition (i.e., full reward, basic reward, and partial reward), and the reward efficiency was measured using BMI changes in children. The results demonstrated that appropriate rewards can motivate children and reduce BMI by encouraging physical activity. Additionally, a discussion that includes the limitation and future work is presented. In the future, the proposed approach will be enhanced to overcome the discussed limitations.

\bibliographystyle{IEEEtran}
\bibliography{ref}



\end{document}